\documentclass[10pt]{article}
\usepackage[dvips]{graphics}

\begin{document}

	\begin{center}
	\LARGE{The masses of the mesons and baryons. \\Part IV. Integer multiple 
rule extension}
	\bigskip

	\Large{E. L. Koschmieder}
	\medskip

	\small{Center for Statistical Mechanics, The University of Texas at 
Austin\\Austin, TX 78712, USA\\e-mail:
koschmieder@mail.utexas.edu}
	\smallskip

	\large{November 2000}
	\end{center}

	\bigskip
	\noindent
	\small{It is shown that the empirical rule that the masses of the stable 
mesons and baryons of the $\gamma$-branch are integer multiples of the 
mass of the $\pi^0$ meson 
with a maximal deviation of 3.3\% holds also for the meson and baryon 
resonances, regardless whether the spin of the particles is 0 or 
$\frac{1}{2}$. It 
is also shown that the masses of the particles with weak decay are integer 
multiples of the 
mass of the $\pi^\pm$ mesons times a common factor 0.853.

	\normalsize

	\section{Introduction}
In a previous article [1] we have shown that the masses of the so-called 
stable mesons and baryons of the $\gamma$-branch are integer multiples of 
the $\pi^0$ meson, with a maximal deviation of 3.3\% and an average 
deviation of 0.73\%. The $\gamma$-branch of the stable mesons and baryons 
consists of the particles whose decay is electromagnetic, the 
characteristic example of the $\gamma$-branch is the $\pi^0$ meson which 
decays via $\pi^0 \rightarrow \gamma\gamma$ (98.8\%). The other branch of 
the 
mesons and baryons is the neutrino branch. The weak decay of the particles 
of the $\nu$-branch is accompanied by the emission of neutrinos, the 
characteristic examples are the $\pi^\pm$ mesons which decay via e.g. 
$\pi^+  
\rightarrow \mu^+ + \nu_\mu$ (99.987\%). The discussion in [1] was limited 
to the 
so-called stable particles. We have explained the integer multiple rule 
of the stable mesons and baryons of the $\gamma$-branch with the standing 
wave model proposed in [2,3]. We will now show that the integer multiple 
rule does not only apply to the stable particles, but also to the meson 
and baryons resonances with isospin  I\,$\leq$\,1 and spin 
J\,$\leq$\,$\frac{1}{2}$.

	\section{The spectrum of the $\gamma$-branch particles}

From a theoretical point of view it is clear that the particles with the 
most simple structure are those with I,J\,=\,0,0, but also with 
strangeness 
S\,=\,0 and charm C\,=\,0. All I,J\,=\,0,0 mesons of the $\gamma$-branch, 
whether 
they are stable or resonances, are listed in Table 1, which is based on 
the Particle Physics Summary [4].
The $\pi^0$ meson does not appear in Table 1 
because its isospin I\,=\,1. Also omitted are two $f_0$ resonances with 
I,J\,=\,0,0 
whose mass is uncertain by more than $\pm3\%$. Included in this Table are 
two 
charmed particles, the $\eta_c$ and $\chi_{c_0}$ mesons; bottom particles 
are 
not considered. A least square plot of the masses of the $\gamma$-branch 
mesons with I,J\,=\,0,0 as a function of the integer N is shown in 
Fig.\,1. The 
integer N is the integer number nearest to the actual ratio m/m($\pi^0$). 
The points in Fig.\,1 refer to resonances, but for the point at N\,=\,4 
which 
represents the $\eta$ meson. 
 
	\begin{table}[t] \caption{The $\gamma$-branch mesons with 
I,\,J\,=\,0\,,0} 
 	\begin{tabular}{lllllrccc||ccllll}\\[0.1cm]
& & & particle  &  m/m($\pi^0$) & N  & & &  &  &  & particle & m/m($\pi^0$) & 
N \\[0.5ex]\hline
& & & $\eta$  &  4.0559  &  4 &  &  & & &  &  $\eta(1440)$  & 10.48  & 10  
\\
& & & $\eta'$  &  7.0958  & 7 &  & & & & &  $f_0$(1500)  &  11.135  & 
11 \\
& & & $f_0$  &  7.261 &  7 & & & & & &  $\eta_c$  &  22.076 &  22 \\
& & & $\eta$(1295)  &  9.594 &  10 & & & & & &  $\chi_{c_0}$  &  25.301 &  
25  
 	\end{tabular}
	\end{table}
	\vspace{0.5cm}
	\includegraphics{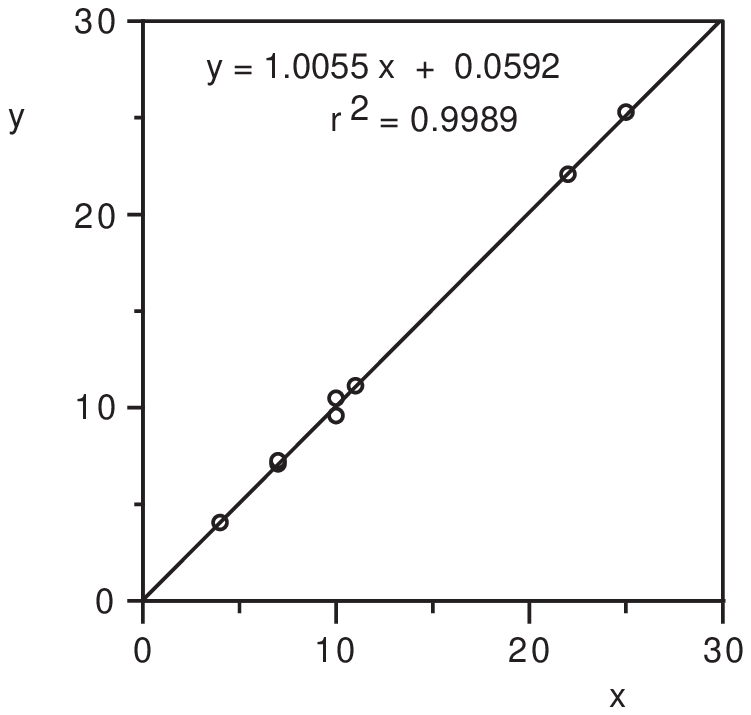}
	\vspace{-0.3cm}
	\begin{quote}  
Fig.1: The mass of the $\gamma$-branch mesons in units of 
m($\pi^0$) as a function of the integer N. y\,=\,m/m($\pi^0$); x is integer N.
	\end{quote}

	\indent
As Fig.\,1 shows the I,J\,=\,0,0 
$\gamma$-branch 
mesons follow the integer multiple rule very well. According to this 
figure the 
masses are determined  by the formula
	\begin{equation}  m(N)/\mathrm{m}(\pi^0) = 1.0055\,N + 0.05923 , 
	\end{equation}
\noindent 
with the almost perfect correlation coefficient 
0.999.
The difference between the line described by (1) and the ideal line for 
integer multiples, which goes through the origin and has slope 1.0000, may 
originate from an improper choice of the reference particle. The 
experimental value of m($\pi^0$) is known to seven decimals, but the 
$\pi^0$ meson is a particle with I\,=\,1, and therefore not necessarily 
the 
proper reference for a line describing the I,J\,=\,0,0 mesons. We do not 
know 
how the value of the isospin affects the mass of a particle. From Eq.\,(1) 
follows 
that m(1)\,=\,1.0647\,m$(\pi^0)$. Presumably this is the mass of a neutral 
meson with I,J\,=\,0,0. It is, by all means, possible that m(1) with 
I,J\,=\,0,0 has a mass 6.5\% larger than m$(\pi^0)$ with I,J\,=\,1,0, or 
whose mass is 3\% larger 
than m$(\pi^\pm)$. The majority of the charged stable particles have 
a mass smaller than the mass of the corresponding neutral particles. From 
the masses of the stable particles it is also apparent that the 
differences between charged and neutral particles is particularly large 
(in percent) in the case of the $\pi$ mesons. According to Eq.\,(1) the 
contribution of the intercept y(0) to the masses of the particles 
decreases with increased N in agreement with the empirical fact that the 
higher N the smaller are, in general, the deviations of the particle 
masses from strictly integer multiples of m$(\pi^0)$.

\indent
Next we look at the masses of the stable baryons and of the baryon 
resonances as a function of N. The masses of the $\gamma$-branch baryons 
with I\,$\leq$\,1,J\,=\,$\frac{1}{2}$ are listed in Table 2. 

	\begin{table}[h] \caption{The $\gamma$-branch baryons with 
I\,$\leq$\,1,J\,=\,$\frac{1}{2}$ }  

	\begin{tabular}{lllr||cllll} \\[0.1cm]
particle & m/m($\pi^0$) & I,\,J  & N & & particle & m/m($\pi^0$) & I,\,J & 
N  
\\[0.5ex]\hline
$\Lambda$ & 8.26577 & 0,\,$\frac{1}{2}$ & 8 & &  $\Xi^0$ & 9.7417 &  
$\frac{1}{2}$,\,$\frac{1}{2}$ &10 \\[0.07cm]
$\Lambda$(1405) & 10.42 &  0,\,$\frac{1}{2}$ & 10 & & $\Lambda^+_c$ & 
16.928 & 
0,\,$\frac{1}{2}$ & 17 \\[0.07cm]
$\Lambda$(1670) & 12.37 & 0,\,$\frac{1}{2}$ & 12 & &  $\Lambda_c$(2593) & 
19.215 &  0,\,$\frac{1}{2}$ & 19  \\[0.07cm] 
$\Lambda$(1800) & 13.33 & 0,\,$\frac{1}{2}$ & 13 & & $\Sigma^0_c$  & 
18.167 & 
1,\,$\frac{1}{2}$ & 18 \\[0.07cm]
$\Sigma^0$ & 8.835 & 1,\,$\frac{1}{2}$ & 9 & & $\Xi^0_c$ & 18.302 & 
$\frac{1}{2}$,\,$\frac{1}{2}$ &18 \\[0.07cm]
$\Sigma$(1660)  & 12.298 & 1,\,$\frac{1}{2}$  & 12  & &  $\Omega^0_c$  & 
20.03 & 0,\,$\frac{1}{2}$ & 20 \\[0.07cm]
$\Sigma$(1750)  & 12.965  & 1,\,$\frac{1}{2}$ & 13 \\
	\end{tabular}
	\end{table}

\noindent
The $\Omega ^-$ particle has been omitted from the list because it has 
spin $\frac{3}{2}$. Including however 
$\Omega^-$ in the list does not alter the following results significantly. 
Also omitted in Table 2 is the $\Lambda$(1810) resonance which differs 
from $\Lambda$(1800) only in parity, and two $\Xi$ resonances whose spin 
is uncertain. A least square analysis of the masses in Table 2 yields 
the formula

	\begin{equation}  m(N)/\mathrm{m}(\pi^0) = 1.0013\,N +\,0.1259 , 
	\end{equation} 

\noindent 
with the very good correlation coefficient 0.997. The baryons and baryon 
resonances follow the integer multiple rule in a good approximation. The 
comparatively large intercept in Eq.\,(2) is of less concern than in the 
case of the mesons because the first mass affected by y(0) is that of the 
$\Lambda$ baryon with N\,=\,8, which means that y(0) contributes only 
1.57\% to 
m$(\Lambda)$. To assess the significance of the difference between 
m$(\Lambda)$ according to Eq.\,(2) and the measured m$(\Lambda)$ we must 
keep in mind that the $\Lambda$ baryon has spin $\frac{1}{2}$ and 
strangeness 
S\,=\,$-$1, 
whereas spin and strangeness of the reference particle $\pi^0$ are both 
zero.
If we combine all mesons and baryons of the $\gamma$-branch with 
I\,$\leq$\,1,J\,$\leq$\,$\frac{1}{2}$ of Tables 1,2 we arrive at Fig.\,2 
which shows that these 22 particles, eight 
I,J\,=\,0,0 mesons, thirteen J\,=\,$\frac{1}{2}$ baryons and the $\pi^0$ 
meson with I,J\,=\,1,0 
follow the integer multiple rule in a good approximation with the nearly 
perfect correlation coefficient 0.999. It is

	\begin{equation}  m(N)/\mathrm{m}(\pi^0) = 1.0056\,N + 0.061026 ,  
	\end{equation} 

\noindent      
which differs only marginally from Eq.\,(1). This is so although the line 
on 
Fig.\,2 describes stable and unstable particles with different spin, 
J\,=\,0 for 
the mesons and J\,=\,$\frac{1}{2}$ for the baryons. Spin $\frac{1}{2}$ 
should make a contribution 
to the energy of the particles but, as Fig.\,2 shows and as we pointed out 
in [1], spin $\frac{1}{2}$ does not alter the integer multiple rule. So do 
different 
values of the isospin or different values of the strangeness, in 
particular S\,$\neq$\,0 for $\Lambda$, $\Sigma$, $\Xi$, and so do 
different 
values of C\,$\neq$\,0 of the charmed baryons and the $\eta_c$ and the 
$\chi_{c_0}$ mesons as well. It is astounding that the integer multiple 
rule 
holds in spite of the differences in the four parameters involved and 
regardless of whether the particles are stable or resonances.

	\vspace{0.5cm}
	\includegraphics{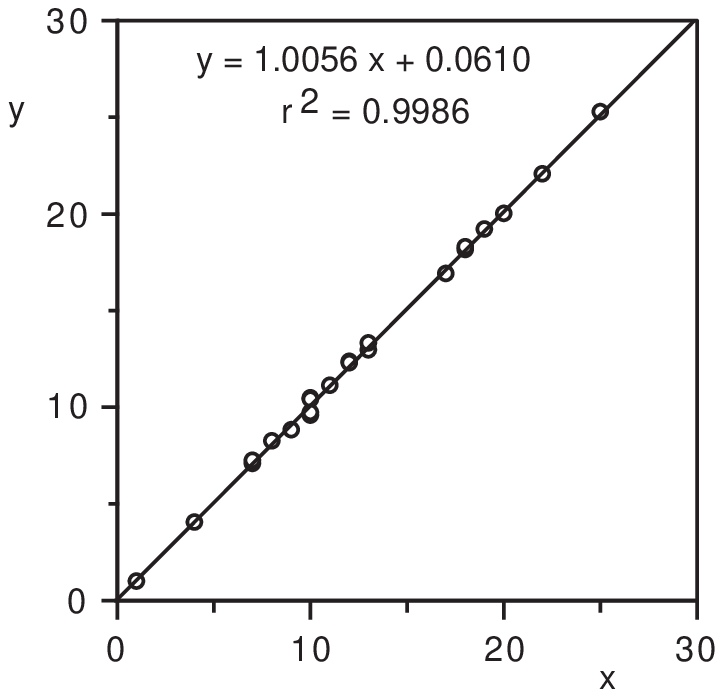}
	\vspace{-0.3cm}
	\begin{quote}  
Fig. 2: The mass of all mesons and baryons with 
I\,$\leq$\,1,J\,$\leq\frac{1}{2}$ in units of m($\pi^0$) as a function of 
the integer N. y = m/m($\pi^0$) ; x is integer N.
	\end{quote}

	\section{The spectrum of the $\nu$-branch particles}

So far we have been concerned with the $\gamma$-branch of the particles. 
The neutrino branch is clearly distinguished from the $\gamma$-branch by 
the neutrinos which are emitted when the particles of the $\nu$-branch 
decay. The particles of the $\nu$-branch, according to the Particle 
Physics Summary, are listed in Table 3 together with the neutron 
resonances.

	\begin{table}[h]\caption{The particles of the $\nu$-branch}
	\begin{tabular}{lllrc||llll} \\[0.1cm]
 
particle &  m/m($\pi^\pm$) & I,\,J & N & & particle & m/m($\pi^\pm$) & 
I,\,J & N \\[0.5ex]\hline
$\pi^\pm$ & 1.0000 & 1,\,0 & 1& &  N(1535) & 10.998 & 
$\frac{1}{2}$,\,$\frac{1}{2}$ & 13 \\[0.07cm]
K$^\pm$ & 3.53713 & $\frac{1}{2}$,\,0 & 4 & &  N(1650) & 11.822 & 
$\frac{1}{2}$,\,$\frac{1}{2}$ &  14 \\[0.07cm]
n  & 6.73186 & $\frac{1}{2}$,\,$\frac{1}{2}$ & 8 & &  D$^\pm$ & 13.393 & 
$\frac{1}{2}$,\,0 & 16  \\[0.07cm]
N(1440) & 10.317 & $\frac{1}{2}$,\,$\frac{1}{2}$ &12 & & D$^\pm_S$ & 
14.104 & 
0,\,0 & 16  \\[0.07cm]

	\end{tabular}
	\end{table}

The reference particles for the $\nu$-branch appear to be the 
$\pi^\pm$ mesons because they are typical for the weak $\nu$-branch 
decays. 
As we have pointed out in [1] the ratios of the masses of the $\nu$-branch 
particles to the mass of the $\pi^\pm$ mesons are integer numbers N 
times a common factor $0.861\,\pm\,0.022$. The ratios cannot be straight 
integer 
multiples of m($\pi^\pm$) because 
m\,(K$^\pm)\,/\mathrm{m}\,(\pi^\pm)=0.8843 \cdot 4=3.537$. 

	\vspace{0.5cm}
	\includegraphics{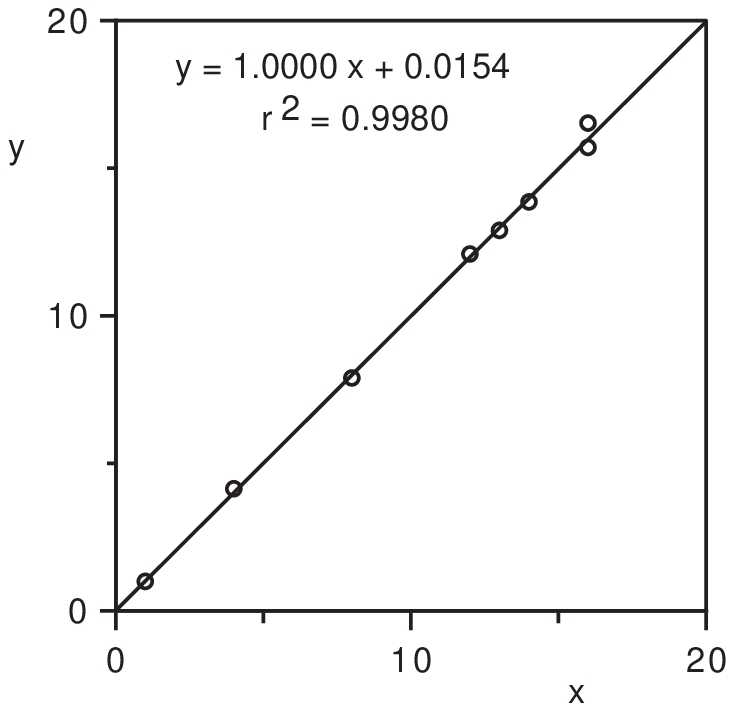}
	\vspace{-0.3cm}
	\begin{quote}   
	Fig. 3: The mass of the neutrino branch particles in units of m($\pi^\pm$) as a function of the integer N. y = m/0.853\,m($\pi^\pm$); x is integer N.
	\end{quote}

\noindent
A least square 
analysis of the 
masses in Table 3 with $m(N)$/0.861\,m\,$(\pi^\pm)$ versus N does, 
however, not have 
a line with slope 1. But a plot of $m(N)$/0.853\,m\,$(\pi^\pm)$ versus N 
(N\,$>$\,1) has 
slope 1.0000 and a small intercept y(0) as shown on Fig.\,3. 

\noindent
The masses of 
the neutrino branch are therefore given by

	\pagebreak
	\begin{equation}  m(N)/0.853\,\mathrm{m}(\pi^\pm) = 1.0000\,N + 0.01537, 
\qquad\mbox{$N\,>\,1$} ,  
	\end{equation}

\noindent 
with the correlation coefficient r$^2$=0.998. If we drop the neutron 
resonances in Table 3 we find
	\begin{displaymath}  m(N)/0.853\,\mathrm{m}(\pi^\pm) = 1.0055\,N + 
0.00575, \qquad\mbox{$N\,>\,1$} .               
	\end{displaymath}

\noindent 
The slope 1.0055 in this formula can, of course, be reduced to 1.0000 by 
proper choice of 
the factor before m$(\pi^\pm)$, i.e. by 0.8574. The common 
factor 0.853 in Eq.\,(4) differs by only 1\% from the empirical value 
0.861 
given in [1], well within the uncertainty $\pm$0.022 of 0.861 given there. 
The 
mass m(1) of Eq.\,(4) is, by definition, the same as m$(\pi^\pm)$, the 
mass 
m(4) is 0.969\,$\cdot$\,m(K$^\pm$) and m(8) is 0.984$\cdot$\,m(n). The 
3.1\% 
difference of m(4) from m(K$^\pm)$ may be the result of the strangeness 
S\,=\,$\pm$1 and of I\,=\,$\frac{1}{2}$ of the K mesons, whereas the 
reference particle
$\pi^\pm$ has S\,=\,0 and I\,=\,1, in other words the $\pi^\pm$ mesons are 
not the perfect 
reference for the K$^\pm$ mesons.

Surprisingly the $\pi^0$ meson provides also a good reference for the 
$\nu$-branch particles. A least square analysis of the eight $\nu$-branch 
particles of Table 3 gives the line
	\begin{equation}  m(N)/\mathrm{m}(\pi^0) = 1.0000\,N - 2.284\cdot10^{-4} ,   
	\end{equation}

\noindent 
with the still very good correlation coefficient r$^2$=0.996. Equation 
(5) makes it appear that the particles of the $\nu$-branch are integer 
multiples of m($\pi^0$). However, the mass of m(4) according to Eq.\,(5) 
is 
1.0936\,$\cdot$\,m(K$^\pm$), which is a much worse fit to m(K$^\pm)$ than 
m(4) 
according to Eq.\,(4), whose m(4) = 0.968$\cdot$m($K^\pm$). On the other hand m(7) according to Eq.\,(5) is 
only 
0.57\% larger than m(n). But neither the K$^\pm$ mesons nor the neutron 
decay 
electromagnetically as the reference $\pi^0$ meson does. Rather both 
particles decay with the emission of neutrinos, K$^\pm$ in 71.5\% of the 
cases and the neutron in 100\% of the cases, and we believe that the 
decays 
tell what the particles are made of. Actually the place N\,=\,4 in 
Eq.\,(5) or 
the place for a particle with four times the mass of the $\pi^0$ meson is 
filled with the $\eta$ meson with the measured 
m$(\eta)=1.014\cdot4\,\mathrm{m}(\pi^0)$. The 
$\eta$ meson however is, as its decay shows, part of the $\gamma$-branch 
of the particles. Therefore Eq.\,(4) describes correctly the neutrino 
branch 
and not Eq.\,(5). The factor 0.853 in front of m$(\pi^\pm)$ in Eq.\,(4) 
will be 
explained in a forthcoming paper as a consequence of the oscillations of a 
neutrino lattice.

	\section{Particles with spin 1}

From the foregoing it might appear that the masses of the particles are 
always integer multiples of a particular reference particle, or integer 
multiples of the mass of a reference particle times a constant factor. 
However the meson resonances with spin 1 tell that this is not necessarily 
so. The meson resonances with spin 1 and S,C\,=\,0,0 according to [4] are 
listed in Table 4.
	\begin{table}[t]\caption{Meson resonances with spin 1}
	\begin{tabular}{llllc||llll} \\[0.1cm]
particle & m/m($\pi^0$) & I,\,J & N & &  particle & m/m($\pi^0$) & I,\,J & 
N \\[0.5ex]\hline
$\rho$(770) & 5.694 & 1,\,1 & 6 & &   f$_1$(1420) & 10.571 & 0,\,1 & 10 \\
$\omega^0$(782) & 5.7932 & 0,\,1 & 6 & &   $\omega$(1420) &10.513 & 0,\,1 
& 10 \\
$\phi^0$(1020) & 7.55253 & 0,\,1 & 8 & &   $\rho$(1450) & 10.854 & 1,\,1 & 
11 \\
h$_1$(1170) & 8.668 & 0,\,1 & 9 & &  f$_1$(1510) & 11.202 & 0,\,1 & 11 \\
b$_1$(1235) & 9.1201 & 1,\,1 & 9 & &  $\omega$(1600) & 12.217 & 0,\,1 & 12 
\\
a$_1$(1230) & 9.1127 & 1,\,1 & 9  & &   $\phi$(1680) & 12.447 & 0,\,1 & 12 
\\
f$_1$(1285) & 9.4994 & 0,\,1 & 9 & &  $\rho$(1700) & 12.595 & 1,\,1 & 13 \\
	\end{tabular}
	\end{table} 
This table contains the pair b$_1$ and a$_1$ as well as the pair 
f$_1$(1420) and 
$\omega$(1420) whose particles differ from another in their 
parities, but in each pair the masses of the particles are nearly the 
same. Considering in the least square analysis only one mass of each pair 
increases the slope and the 
intercept in Eq.\,(6) substantially. 
A least square analysis of the 14 particles in Table 4 yields a straight 
line 
given by 
	\begin{equation} m(N)/\mathrm{m}(\pi^0) = 1.0257\,N - 0.3345, 
\qquad\mbox{r$^2$=0.977.} 
	\end{equation} 

\noindent 
The intercept y(0) in (6) is much larger than the intercepts in Figs.\,1,2 
and the difference of the correlation coefficient from a perfect 1.0000 is 
by an order of magnitude larger than in Figs.\,1,2. It seems to be 
questionable that the masses of the meson resonances with spin 1 can be 
described satisfactorily by an integer multiple rule.

	\section{Conclusions}

We have found that the masses of the stable mesons and of the 
meson resonances of the $\gamma$-branch are a linear function of the 
variable N which denotes integer multiples of the mass of the $\pi^0$ 
meson. The correlation between m(N) and N has a nearly perfect correlation 
coefficient r$^2$=0.999. The same applies when the mesons and meson 
resonances with I$\le$1,J\,=\,0 are combined with the baryons and baryon 
resonances with I$\le$1,J\,=\,$\frac{1}{2}$. Spin $\frac{1}{2}$ does not 
seem to affect the integer 
multiple rule, neither does strangeness nor charm. We have, furthermore, 
confirmed our previous finding that the masses of the particles with weak 
decay, i.e. of the $\nu$-branch, are integer multiples of the $\pi^\pm$ 
mesons times a common factor. This rule holds with the near perfect 
correlation coefficient r$^2$\,=\,0.998.

	\bigskip

	\noindent
	\textbf{REFERENCES}
\smallskip

[1] E.L.Koschmieder, xxx.lanl.gov/abs/hep-ph/0002179.\\
\hspace*{3em}Bull. Acad. Roy. Belgique,\,{\bfseries X},\,281 (2000).
\smallskip

[2] E.L.Koschmieder and T.H.Koschmieder,\\
\hspace*{3em}xxx.lanl.gov/abs/hep-lat/0002016.\\
\hspace*{3em}Bull. Acad. Roy. Belgique,\,{\bfseries X},\,289 (2000).
\smallskip

[3] E.L.Koschmieder, xxx.lanl.gov/abs/hep-lat/0005027.
\smallskip

[4] R.Barnett et al.,\,Rev.Mod.Phys.\,{\bfseries 68},\,611 (1996).

	\end{document}